\documentclass[preprint,showpacs,amsmath,amssymb,pre,superscriptaddress,floatfix]{revtex4-1}

\usepackage{graphicx}% Include figure files
\usepackage{dcolumn}% Align table columns on decimal point
\usepackage{bm}% bold math
\usepackage{amssymb}
\usepackage{multirow}
\usepackage{braket}

\begin{document}

\title{Diffusion and transport in locally disordered driven lattices. }

\author{Thomas Wulf}
    \email{Thomas.Wulf@physnet.uni-hamburg.de}
    \affiliation{Zentrum f\"ur Optische Quantentechnologien, Universit\"at Hamburg, Luruper Chaussee 149, 22761 Hamburg, Germany}

\author{Alexander Okupnik}
    \affiliation{Zentrum f\"ur Optische Quantentechnologien, Universit\"at Hamburg, Luruper Chaussee 149, 22761 Hamburg, Germany}

\author{Peter Schmelcher}
    \email{Peter.Schmelcher@physnet.uni-hamburg.de}
    \affiliation{Zentrum f\"ur Optische Quantentechnologien, Universit\"at Hamburg, Luruper Chaussee 149, 22761 Hamburg, Germany}
    \affiliation{The Hamburg Centre for Ultrafast Imaging, Universit\"at Hamburg, Luruper Chaussee 149, 22761 Hamburg, Germany} 

\date{\today}


\begin{abstract}

We study the effect of disorder on the particle density evolution in a classical Hamiltonian driven lattice setup.
If the disorder is localized within a finite sub-domain of the lattice, the emergence of strong tails in the density distribution which even increases towards larger positions is shown, thus yielding a highly non Gaussian particle density evolution.
As the key underlying mechanism we identify the conversion between different components of the unperturbed systems mixed phase space which is induced by the disorder.
Based on the introduction of individual conversion rates between chaotic and regular components, a theoretical model is developed which correctly predicts the scaling of the particle density.
The effect of disorder on the transport properties is studied where a significant enhancement of the transport for cases of localized disorder is shown, thereby contrasting strongly the merely weak modification of 
transport for global disorder.

% We study the impact of disorder in a classical Hamiltonian driven lattice setup.
% Collisions with the disorder barriers lead to conversion of particles between regular and chaotic components of 
% the phase space of the unperturbed lattice. If the disorder is localized to a finite domain, particles are accumulated 
% into the regular structures, which is shown to lead to a highly non Gaussian particle density evolution. The obtained 
% densities are reproduced to excellent agreement by means of a theoretical model based on a decomposition of the particle ensemble into multiple regular and one diffusive part.   
% Finally, we study the effect of disorder on the transport properties where we provide an extension to the sum rule of Hamiltonian ratchet transport which incorporates the conversion of particles into regular structures as induced by the disorder.

\end{abstract}

\maketitle

\section{Introduction}

Driven lattice setups have long been one of the paradigmatic examples for the study of complex, out of equilibrium systems.
In particular, their capability to transport particle ensembles in a controlled manner, even though all applied forces are zero mean valued, has stimulated considerable amounts of research and has under the label 'ratchet physics' found direct applications in various physical settings \cite{schanz_classical_2001,schanz_directed_2005,serreli_molecular_2007,denisov_tunable_2014,liebchen_interaction_2015,reimann_brownian_2002,hanggi_artificial_2009}. 
Noteworthy examples are Brownian or molecular motors \cite{astumian_fluctuation_1994,astumian_thermodynamics_1997,julicher_modeling_1997,reimann_brownian_2002,hanggi_artificial_2009,schmitt_molecular_2015,roth_optimization_2015},
or the controlled separation of particles or even living bacteria of different species \cite{matthias_asymmetric_2003,savelev_separating_2005,bogunovic_particle_2012,tahir_dynamically-tunable_2014}.
Since the uprising of cold atom physics, experiments have also been performed on atoms exposed to time dependently modulated optical standing waves where the controlled locomotion of atomic ensembles was realized 
\cite{mennerat-robilliard_ratchet_1999, gommers_quasiperiodically_2006, schiavoni_phase_2003, wickenbrock_vibrational_2012}.
Here, at temperatures well above the Bose Einstein transition temperature, it was shown how the atoms dynamics can be described reliably by means of a purely classical description \cite{wickenbrock_vibrational_2012}. 
Whereas most of the original works on ratchet physics are focused on the strongly damped or even over damped situation,
in cold atom experiments, one could conveniently reach the Hamiltonian, that is the dissipation free, limit \cite{wickenbrock_vibrational_2012}. 
Another promising experimental realization of ratchets, which allow for a precise control over the systems parameters and thus are ideally suited to study the dissipation-less regime, are SQUIDs \cite{weiss_ratchet_2000} where the phase of the macroscopic wave function is indeed described by a classical equation of motion \cite{zapata_voltage_1996,spiechowicz_josephson_2014}.
From a theoretical viewpoint, Hamiltonian ratchets have been analyzed comprehensively \cite{flach_directed_2000, schanz_classical_2001, schanz_directed_2005} and e.g. a sum rule was derived by which the systems asymptotic transport velocity can be deduced directly from the composition of the underlying mixed phase space \cite{schanz_classical_2001}.

Thereby, at the very heart of the analysis of Hamiltonian driven lattices, including the mentioned sum rule, is the apparent concept that different chaotic or regular components of the phase space are not mixed under time evolution.
Recently however, setups have been proposed which allow for a conversion of particles from the systems chaotic layer into its regular components and vice versa 
\cite{wulf_disorder_2014,liebchen_interaction-induced_2012,petri_formation_2011,wulf_analysis_2012}.
In \cite{wulf_disorder_2014} this was achieved by introducing randomly placed impurities into an otherwise translationally symmetric arrangement of scatterers. Here, the collisions with the impurities were shown to induce 
hopping processes between different components of the unperturbed systems phase space. Even more, if the occurrence of impurities was restricted to a finite sized sub-domain of the lattice,
an accumulation of particles into the regular structures of the unperturbed systems phase space was reported.
Similarly, it was shown in \cite{liebchen_interaction-induced_2012} how weak inter particle interactions again cause a conversion of particles into the regular structures of the phase space of the non interacting system.
Here, it was also demonstrated that this subsequent filling of regular structures alter the systems transport properties, even leading to time dependent reversals of the transport direction.
In a very different setting, it was also shown in \cite{petri_formation_2011, wulf_analysis_2012} how a block lattice which is build up out of different domains allows for transitions 
between different components of the phase spaces of the individual blocks.

% In REF this was achieved by means of inter particle collisions, where the subsequent filling of the phase space's regular structures was shown to significantly alter the systems transport properties, even leading to time dependent reversals of the transport direction.
% Similarly, it was shown in REF how randomly placed impurities in a lattice lead to a comparable effect.  Specifically, REF demonstrated how collisions with the impurities even lead to an accumulation of particles in the regular structures of the unperturbed, i.e. impurity free, systems phase space, if the impurities are located only within a finite sub-domain of the entire lattice.
It is quite apparent that such conversion between the chaotic layer and regular structures lead to a significant modification of usual diffusive properties inherit to motion purely within the chaotic layer. However, a systematic investigation of the diffusive properties, or more generally speaking of the characteristics of the spatial particle density evolution, for systems with an inherent conversion between different components of a Hamiltonian mixed phase space is still missing, and is indeed the main subject of the present manuscript. 
To this end, we investigate a Hamiltonian system consisting of a temporally shaken lattice potential with additional randomly placed Gaussian barriers, acting as impurities. Similar to the results obtained in \cite{wulf_disorder_2014}, we show that, by means of collisions with the impurity barriers, particles undergo conversion processes between different components of the mixed phase space of the unperturbed lattice. In cases were the occurrence of impurities is restricted to a finite domain of the lattice, we show that the spatial particle density develops strong tails with even increasing density towards larger positions. 
We provide a detailed analysis of the observed density evolution and propose a theoretical model, based on a decomposition of the entire particle ensemble into regular and diffusive sub-ensembles,
which correctly predicts the scaling of the tails of the density.
Furthermore, we argue how the inclusion of a localized disorder domain in a Hamiltonian ratchet may significantly enhance the directed current of the setup. Here, we give a modified version of the established sum rule of Hamiltonian ratchet transport \cite{schanz_classical_2001} which incorporates the conversion from the chaotic layer into the regular structures of the unperturbed lattice. 

Our manuscript is structured in the following way:
In Sec.\,\ref{S1} we introduce the setup of a disordered lattice. Section \ref{S2} sums up 
the main results of disorder induced regular motion as described in \cite{wulf_disorder_2014}. In Sec.\,\ref{S3} 
we investigate the particle density evolution in a locally disordered lattice and derive 
a model which reproduces the correct scaling of the densities tails. Section \ref{S4} is devoted 
to the transport properties of the lattice and how they are influenced by the disorder. Finally, we conclude in Sec.\,\ref{S5}.

\section{The disordered driven lattice.}
\label{S1}

\begin{figure}[htbp]
\centering
\includegraphics[width=0.5\columnwidth]{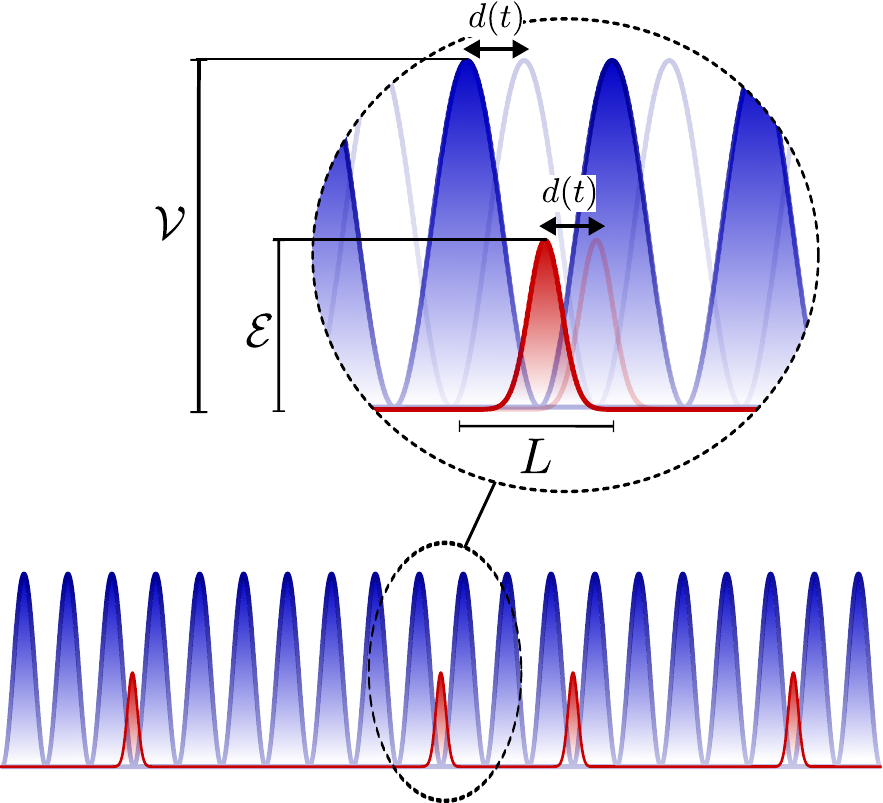}
\caption{\label{fig:setup}Snapshot of the lattice including the impurities in the form of Gaussian barriers which are placed randomly in the minima of the lattice. 
In the zoom, the lateral shaking of the entire system according to the driving law $d(t)$ is indicated together with the most relevant system parameters: the potential height of the lattice $\mathcal{V}$ and of the 
impurities $\mathcal{E}$ as well as the lattice spacing $L$. }
\end{figure}

We consider classical, non interacting particles of mass $m$ in one spatial dimension. 
Their dynamics is governed by the time-dependent Hamiltonian $H(x,p,t)=\frac{p^2}{2m}+V(x,t)$ with time $t$, position $x$, momentum $p$
and potential $V(x,t)$. 
Thereby, the particles are
subjected to a laterally driven lattice potential, which is perturbed by randomly placed Gaussian barriers. That is the total potential $V(x,t)$ is given as the sum
of a spatially periodic lattice potential, $V_L(x,t)$, with $V_L(x,t)=V_L(x+L,t)$ and a disorder potential $V_D(x,t)$. 
Specifically, the lattice potential $V_L(x,t)$ as well as the randomized term $V_D(x,t)$ are given by:
\begin{equation}
\begin{aligned}
  V(x,t) &= V_L(x,t)+V_D(x,t) \\
  &= \mathcal{V} \cos^2[k(x - d(t))] + \mathcal{E}\sum_{n=-N_D}^{N_D} \chi_n \exp\left[-\left(\frac{x-d(t)-nL}{\sigma} \right)^2\right],
  \label{eq:potential}
\end{aligned}
\end{equation}
where $k\equiv \pi/L$ is the wave vector, $\mathcal{V}$ and $\mathcal{E}$ denote the strength of the lattice and the disorder potential and $\sigma$ is the Gaussian's widths. $\chi_n$
is a randomized sequence of zeros and ones where we take the probability for a given element of $\chi_n$ to be one as $10\%$ and finally, $N_D$ fixes the extension of the disorder domain to $2N_DL$.
Both the lattice and the perturbations are laterally driven according to the driving law:
\begin{equation}
 d(t)=A \cos(\omega t)
 \label{eq:driving}
\end{equation}
with driving amplitude $A$ and frequency $\omega$ and hence, the Gaussians are centered in the minima of the lattice at all times
(see Fig.\,\ref{fig:setup} for an illustrative sketch of the system).
It should be noted that, by transforming into the co-moving frame $x\rightarrow \tilde x= x-d(t)$, one can show straightforwardly that the transformed Hamiltonian $\tilde H$ describes a particle exposed 
to a static lattice which is of the form of $V(x,t)$ with $d(t)=0$ and an additional spatially constant oscillating force term $\propto \cos(\omega t)$. 

In the following, we will investigate and compare three different scenarios: first, the unperturbed case with $\mathcal{E}=0$ and thus $V=V_L$. 
% Second, the case with a Gaussian in every potential minimum, i.e. we set $\chi_n = 1,\ \forall n$, which again corresponds to an ``unperturbed'' lattice in the sence that spatial periodicity remains intact.
Second, the case of a \textit{globally} disordered lattice as provided by Eq.\,(\ref{eq:potential}) for $N_D\rightarrow \infty$
and finally, and this will actually be the main focus throughout this work, we will study the effects of a \textit{localized} disorder region as realized by a finite, nonzero value of $N_D$.
Quantities, such as the spatial or momentum densities will be labeled with superscripts ``$0$'', ``$GD$''
or ``$LD$'' for the cases of no disorder, global disorder or local disorder respectively.

%%%%%%%%%%%%%%%%%%%%%%% FIG 1

%%%%%%%%%%%%%%%%%%%%%%%%%%%%%%%%%%%%%%%%%%%%%%%%%%%%%%%%%%%%%%%%%%%%%%%%%%%%%%%%%%%%%%%%%%%%%%%%%%%%%%%%%%%%%%%%%%%%%%%%%%%%%%%%%%%%%%%%%%%%%%%%%%%%%%%%%%%%%%%%%%%%%%%%%%%%%%%%%%%%%%%%%%%%%%%%%
\section{Unperturbed dynamics and disorder induced entering of regular structures.}
\label{S2}
\begin{figure}[htbp]
\centering
\includegraphics[width=0.5\columnwidth]{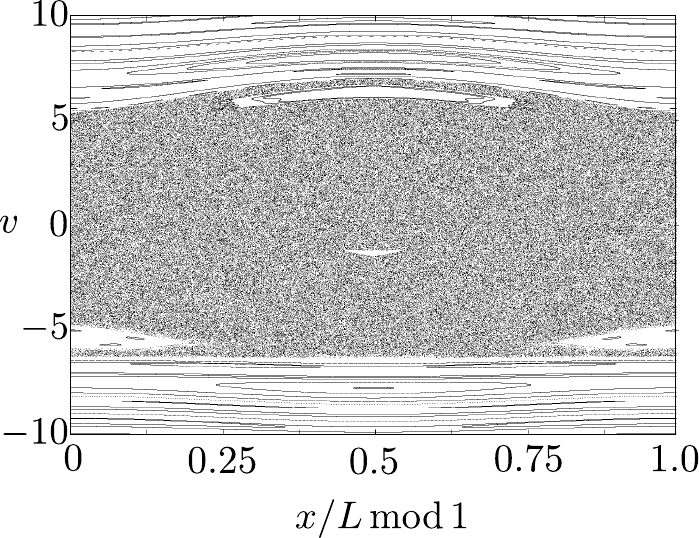}
\caption{\label{fig:PSUS} Poincar\'{e} section of the unperturbed system, that is for $\mathcal{E}=0$ in Eq.\,(\ref{eq:potential}). Remaining parameters are $\mathcal{V}=4$, $k=0.79$, $\omega=3$ and $A=1$. 
Recording velocities and positions at different phases of the driving would reveal that the Poincar\'{e} section respects indeed the symmetry of the Hamiltonian of $x\rightarrow -x$, $v\rightarrow -v$ and $t\rightarrow t + \pi/\omega$.}
\end{figure}

\begin{figure}[htbp]
\centering
\includegraphics[width=0.5\columnwidth]{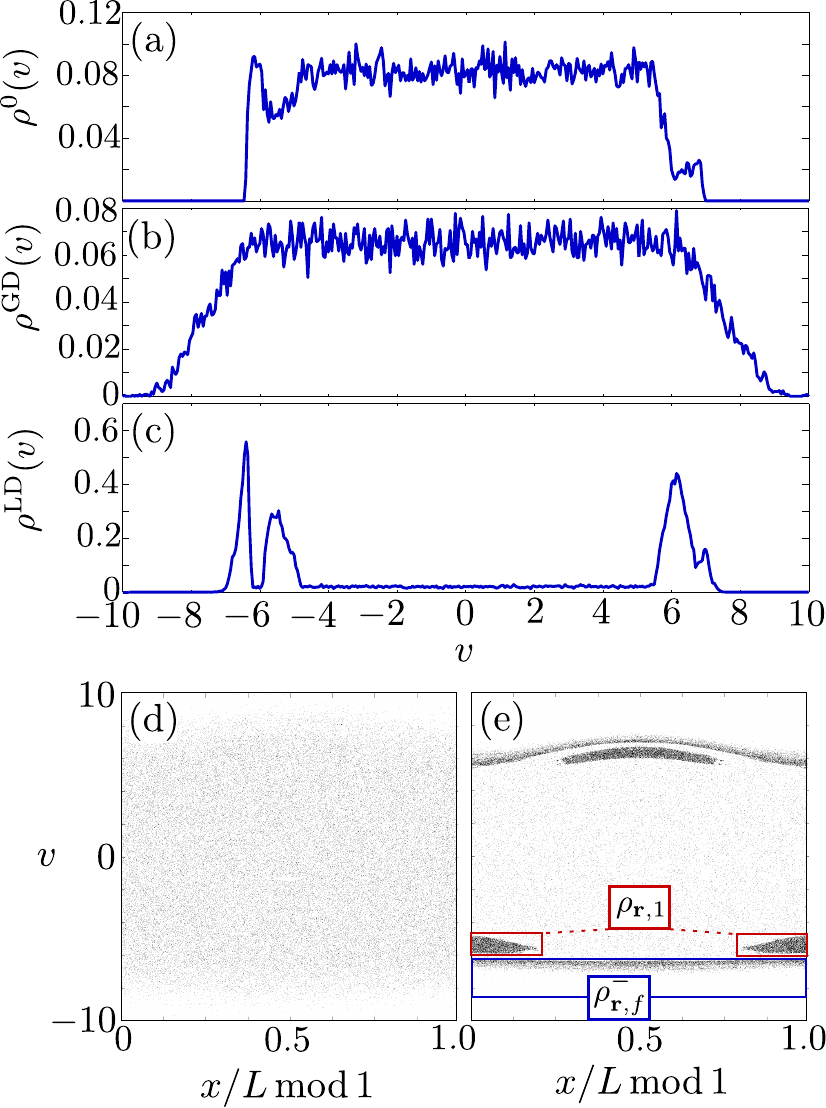}
\caption{\label{fig:rho_v} Velocity distributions $\rho(v)$ at time $t=10^5\times T$ for no disorder (a), global disorder (b) and local disorder with $N_D=5000$ (c). 
Phase space distribution at time $t=10^5\times T$ for global (d) and local (e) disorder.
Disorder 
strength and width are $\mathcal{E}/\mathcal{V}=0.5$ and $\sigma/L=0.3535$; remaining parameters as in Fig.\,\ref{fig:PSUS}.}
\end{figure}

Lets us very briefly recap the main features of the dynamics of a periodically driven, spatially periodic lattice, that is, the case for $ \mathcal{E}=0$ in Eq.\,(\ref{eq:potential}). 
Such a setup features a mixed phase space which can be visualized by a stroboscopic Poincar\'{e} surface of section (PSS). Here, the particles positions and momenta are 
recorded at times $0,T,2T,...$ with $T=2\pi/\omega$ being the temporal period. Furthermore, due to the systems spatial periodicity, the positions can be mapped back to the first unit cell.
The PSS with the parameters as used throughout this work is shown in Fig.\,\ref{fig:PSUS} and reveals the typical ingredients of a mixed phase space \cite{lichtenberg_regular_1992}: a chaotic layer, ballistic islands embedded in it and 
invariant curves at higher energies. Of particular importance for what is to follow, are the two first invariant curves, which span the entire PSS and thus contain the chaotic layer towards higher and lower momenta, respectively. 
These will be referred to as FISCs in the following (First Invariant Surface spanning Curves).
For a particle ensemble initialized at small velocities, the particles fill the chaotic layer ergodically, and the ensembles velocity distribution $\rho^0(v,t)$ is limited by the velocities corresponding to the motion on the two FISCs. 
Furthermore the ballistic islands constitute ``forbidden'' sub-manifolds of the phase space, leading to localized minima of $\rho^0(v,t)$ at velocities where the PSS features ballistic islands (see Fig.\,\ref{fig:rho_v} (a)). 

%%%%%%%%%%%%%%%%%%%%%%% FIG 2
%%%%%%%%%%%%%%%%%%%%%%%%%%%%%%%%%%%%%%%%%%%
We will now explain how the particle dynamics is altered by the inclusion of disorder, that is by setting $\mathcal{E} \neq 0$ in Eq.\,(\ref{eq:potential}). In \cite{wulf_disorder_2014} it was demonstrated how 
randomized deviations from an otherwise spatially periodic system allow for a conversion process of particles from the chaotic layer into the regular structures. That is, the particles dynamics 
can still be described correctly by the phase space of the unperturbed system (henceforth referred to as PSUS), 
until a particle collides with one of the lattice impurities. At this point, the particle dynamics is no longer governed by the PSUS 
and a particle initially located within the chaotic layer of the PSUS may leave the scattering region of the impurity with phase space coordinates corresponding to a regular structure, e.g. a ballistic island, in the PSUS.
Hence, such a particle would have undergone a conversion process from diffusive motion corresponding to the chaotic layer to regular motion in the ballistic island, where the opposite process, i.e. conversion from a ballistic islands into the chaotic layer of the PSUS, is equally possible. Indeed, the phase space distribution of an ensemble initially located within the chaotic layer of the PSUS and then evolving in the globally disordered system, reveals 
that the ballistic islands of the PSUS are now populated uniformly (see Fig.\,\ref{fig:rho_v} (d)). 
This effect of a uniform filling of the phase space can also be seen in the velocity distribution $\rho^{GD}(v)$ (see Fig.\,\ref{fig:rho_v} (b)), which 
does not show any substructure, such as local minima around distinct velocities which would hinge at unaccessible regular structures.

%%%%%%%%%%%%%%%%%%%%%%% FIG 3
%%%%%%%%%%%%%%%%%%%%%%%%%%%%%%%%%%%%%%%%%%%
It was also shown in \cite{wulf_disorder_2014} that, in an arrangement of individual scatterers, a localization of the disorder to a certain domain causes the particles to even accumulate in the regular structures of PSUS.
This holds also in our setup, as seen clearly in the phase space distribution for a disorder domain of $10^4$ lattice sites (i.e. $N_D=5000$) as shown in Fig.\,\ref{fig:rho_v} (e)
and also in the corresponding velocity distribution $\rho^{LD}(v)$ (Fig.\,\ref{fig:rho_v} (c)),
thus indicating the universality of the effect of disorder induced regular dynamics as reported in \cite{wulf_disorder_2014}. 
We will now briefly sum up the two main mechanisms leading to this accumulation (a more detailed discussion can be found in \cite{wulf_disorder_2014}). 
Imagine a particle initialized at $x=0$ which leaves the disorder region for the first time. According to the discussion above, when the particle passes one of the edges of the disorder region at $x=\pm X_D\equiv \pm N_D \times L$, its phase space 
coordinates can either correspond to the chaotic layer of the PSUS or to one of its regular structures, i.e. to a ballistic island or one of the invariant curves above/ below the FISC. Since the ballistic islands as well as the invariant curves above/below the FISCs do not cross $v=0$, particles located within one of those structures while leaving the disorder region can never enter it again, and hence remain within the regular structure. 
On the contrary, if the particle crosses $X_D$ while moving within the chaotic layer, it may return to the disorder region after some time, and hence has another chance to be injected into a regular structure by collisions with the disorder barriers. The second mechanism for the accumulation is more subtle and hinges on the distribution of L\'{e}vy flight lengths of particles in different phase space structures. Here, one can show that, even within the disorder region, particles tend to dwell in the regular structures of the PSUS for comparably long times once they enter them. 
While following such a structure, the particles move ballistically through the lattice and hence overcome long distances, in particular much longer distances than they would overcome in the same time by the diffusive motion corresponding to the chaotic layer. It is thus more likely for a particle to reach $X_D$ while being within a regular structure than in the chaotic layer, which again adds to the accumulation effect.

%%%%%%%%%%%%%%%%%%%%%%%%%%%%%%%%%%%%%%%%%%%%%%%%%%%%%%%%%%%%%%%%%%%%%%%%%%%%%%%%%%%%%%%%%%%%%%%%%%%%%%%%%%%%%%%%%%%%%%%%%%%%%%%%%%%%%%%%%%%%%%%%%%%%%%%%%%%%%%%%%%%%%%%%%%%%%%%%%%%%%%%%%%%%%%%%%
\section{Density evolution.}
\label{S3}
For the unperturbed lattice, the spatial density $\rho^0(x,t)$ of an ensemble which ergodically fills the chaotic layer may at least to some approximation be described by a Gaussian distribution whose width increases monotonically in time according to some diffusion coefficient. The density is in fact non Gaussian due to the trajectories tendency to become "sticky" to regular structures and the associated build up of pronounced tails in $\rho^0(x,t)$. However, in particular close to the ensembles mean position a Gaussian remains a reasonable and often applied approximation. As discussed before, a localized disorder region leads to 
a conversion from diffusively moving particles into ballistically moving ones and hence, we can expect that Gaussian approximations of $\rho^{LD}(x,t)$, which are based on the assumption of a diffusive motion, break down entirely. In the following we discuss how the density in such a locally disordered lattice evolves in time. Furthermore we propose an insightful model, which is based on a decomposition of the ensemble into diffusively and regularly moving particles, and which is able to account for the main features in $\rho^{LD}(x,t)$.

\begin{figure}[htbp]
\centering
\includegraphics[width=0.5\columnwidth]{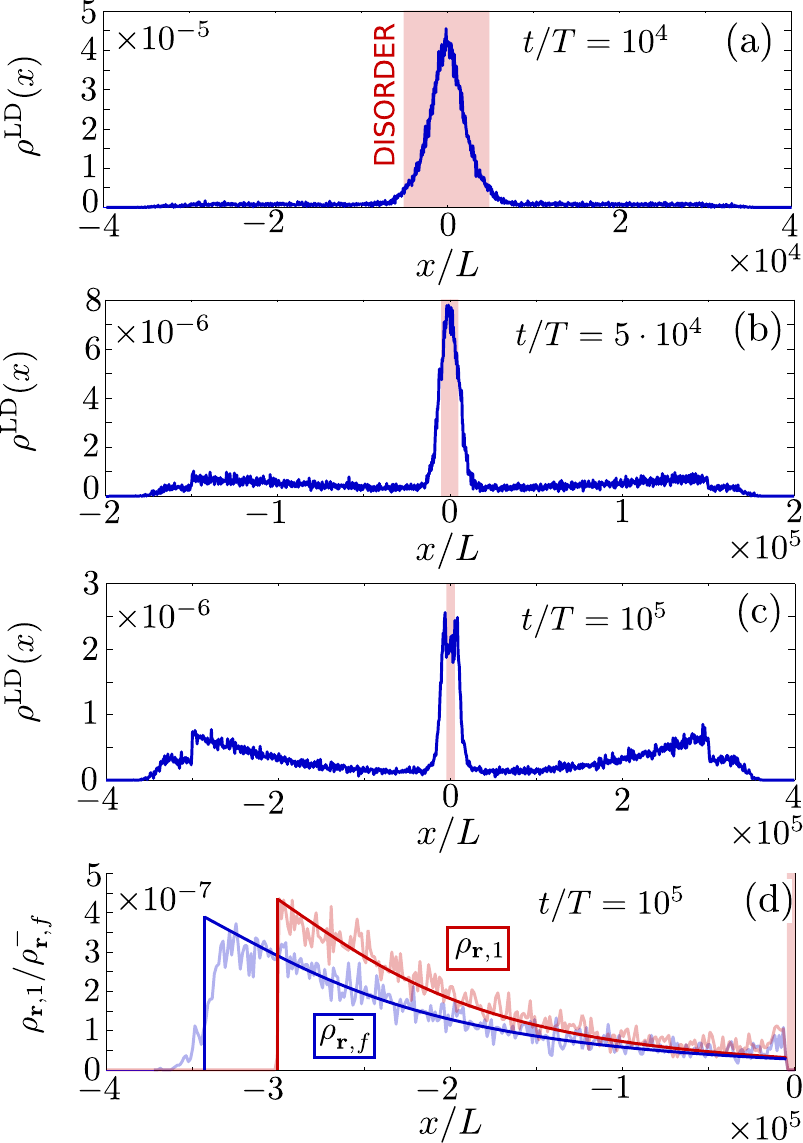}
\caption{\label{fig:rho_x} Density $\rho^{LD}(x,t)$ for a localized disorder domain (red) at times $t/T=10^4$ (a), $5\cdot10^4$ (b) and $10^5$ (c). Parameters as in Fig.\,\ref{fig:rho_v}.
(d) Decomposition of the density according to Eq.\,(\ref{eq:density_sum}) at  $t/T=10^5$ and for $x<0$. 
For $x<<0$, the density of the non regular particles, $\rho_{\textbf{nr}}$ is practically zero such that the entire density in the tails is given as: $\rho^{LD}\approx \rho_{\textbf{r},1} + \rho_{\textbf{r},f}^-$, where
$\rho_{\textbf{r},1}$ contains all particles within the ballistic island at $v\sim -5.5$ of the PSUS and 
$\rho_{\textbf{r},f}^-$ all particles below the FISC at negative velocities (see also Fig.\,\ref{fig:rho_v} (e)). 
The rough, shaded lines denote numerical results while the smooth lines show the densities according Eqs.\,(\ref{eq:conversion_sep}),\,(\ref{eq:density}) with 
no free fit parameters.}
\end{figure}

In Fig.\,\ref{fig:rho_x} we show the temporal evolution of the spatial density for an ensemble of $10^4$ particles with initial velocities $-0.1<v_0<0.1$ and initial 
positions $-10^3\times L < x_0 < 10^3 \times L$ such that all particles are located well within the disorder order region ranging from $-5000\times L$ to $+5000\times L$.
We very clearly see strong deviations from a diffusive, Gaussian spreading of $\rho^{LD}(x,t)$ at larger times. Here, strong tails in the distribution emerge which even increase for larger $x$. Furthermore, a dip in the density within the domain of disorder emerges as seen best in Fig.\,\ref{fig:rho_x} (c). In the following we will show how both the tails in $\rho^{LD}(x,t)$ as well as the depletion within the disorder region can be understood.

%%%%%%%%%%%%%%%%%%%%%%% FIG 4
%%%%%%%%%%%%%%%%%%%%%%%%%%%%%%%%%%%%%%%%%%%%%%%%%%%%%
Above, we argued that particles which leave the disorder region within a regular structure of the PSUS remain within this structure permanently, as they perform no further collisions with the impurity barriers and can thus not be converted back into the chaotic layer. Let us denote the total number of particles which are confined permanently to a regular structure of the PSUS as $N_{\textbf{r}}(t)$. That is, $N_{\textbf{r}}(t)$ includes all particles within a regular structure  of the PSUS and which are located outside of the disorder region. The remaining particles will be denoted as $N_{\textbf{nr}}(t)$, which thus includes all particles within the disorder region and additionally 
particles outside the disorder which are located in the chaotic layer of the PSUS.
The total number of particles $N$ is then given simply by $N=N_{\textbf{r}}(t)+N_{\textbf{nr}}(t)$. The same partition may be employed for the spatial density, hence we write the density of the total ensemble as: 
\begin{equation}
\rho^{LD}(x,t)=\rho_{\textbf{r}}(x,t) + \rho_{\textbf{nr}}(x,t).
\end{equation}
We will confirm later that $\rho_{\textbf{r}}$ describes the tails in $\rho^{LD}$ and we deduce the overall scaling of $\rho_{\textbf{r}}$ in the following. 

By construction, the number of particles which are confined permanently to a regular structure of the PSUS, $N_{\textbf{r}}(t)$, cannot decrease, while particles associated to $N_{\textbf{nr}}(t)$ may be converted into permanently  particles and thus add to $N_{\textbf{r}}(t)$.
This conversion, in the sense of $N_{\textbf{r}}(t)$ and $N_{\textbf{nr}}(t)$ as defined above, may only occur at the instance when a particle leaves the disorder region and thus crosses the positions $\pm X_D$. It seems a reasonable assumption that the number of particles which cross $x=\pm X_D$ per time, and thus also the number of particles which are converted from $N_{\textbf{nr}}(t)$ to $N_{\textbf{r}}(t)$,
is assumed proportional to the number of particles within the disorder region:
\begin{eqnarray}
\frac{\partial}{\partial t} N_{\textbf{r}}(t) \approx \frac{\gamma}{T} \int_{-X_D}^{+X_D} \rho^{LD}(x,t) \, dx
= \frac{\gamma}{T} \int_{-X_D}^{+X_D} \rho_{\textbf{nr}}(x,t) \, dx
%= N-\frac{\partial}{\partial t} N_{\textbf{nr}}(t),
\label{eq:conversion}
\end{eqnarray}
with some proportionality constant $\gamma$ which denotes the fraction of the number of particles within the disorder region which is converted into regular structures 
per driving period $T$
% the ratio of particles which are converted into regular structures per driving period $T$
% denotes the conversion rate into the regular structures per driving period $T$
and the second equality holds because, by definition, $\rho^{LD}=\rho_{\textbf{nr}}$ for $|x|<X_d$.
We will show later, how the temporal evolution of $\partial_t N_{\textbf{r}}(t)$ links directly to the spatial dependence of the tails or the particle density at large positions. Note that
the only time dependence in Eq.\,(\ref{eq:conversion}) is in the integrated density, i.e. in the number of particles which are located within the disorder domain at time $t$: $ \int_{-X_D}^{+X_D} \rho_{\textbf{nr}}(x,t) \, dx$.
In the following we deduce the scaling of this quantity which will then be used to derive the spatial dependence of the density outside the disorder region.

As a "zeroth order" guess, one might assume that the part of the ensemble described by $\rho_{\textbf{nr}}(x,t)$, behaves as if there would be no disorder at all. This however, would neglect the conversion of particles 
into regularly moving ones which are described by $\rho_{\textbf{r}}$ and hence causes a particle loss in $\rho_{\textbf{nr}}$. Lets us now account for this loss effect by indeed assuming, that 
$\rho_{\textbf{nr}}$ is given by $\rho^0$, but with an additional particle loss given by the conversion into regular particles with rate $\gamma$ (cf. Eq.\,\ref{eq:conversion}). The number of particles within the disorder region 
would then be given by:
\begin{equation}
 \int_{-X_D}^{+X_D} \rho^{LD}(x,t)\, dx \approx (1-\gamma)^{t/T} \int_{-X_D}^{+X_D} \rho^{0}(x,t)\, dx,
 \label{eq:rho_ld}
\end{equation}
where the factor $\int_{-X_D}^{+X_D} \rho^{0}(x,t)\, dx$ reflects that we assume that $\rho^{LD}$, within the disorder domain, behaves as if there would be no disorder at all
and the factor $(1-\gamma)^{t/T}$ accounts for the fact that subsequently particles are removed from the disorder domain 
by being permanently converted into regular structures.
% That is, we first assume that $\rho^{LD}$, within the disorder domain, behaves as if there would be no disorder at all, but then take into account that subsequently particles are removed from the disorder domain 
% by being permanently converted into regular structures.
For the density of the unperturbed system, we may furthermore assume a Gaussian: $\rho^0(x,t)= 1/\sqrt(\pi)\sigma_t \exp[-(x/\sigma^0_t)^2] $ where the time dependent width is assumed to be of the form $\sigma^0_t=L\sigma_1(t/T+\sigma_2)^{\sigma_3}$ with three fit parameters $\sigma_{1,2,3}$. 
For the number of particles in the domain $|x|<X_d$ in the unperturbed system, this yields:
\begin{equation}
 \int_{-X_D}^{+X_D} \rho^{0}(x,t) \approx \text{erf}\left(\frac{X_d}{\sigma^0_t}\right).
 \label{eq:rho_0}
\end{equation}
By explicitly simulating $\rho^0(x,t)$ we can justify this approximation and also fix the involved fit parameters $\sigma_{1,2,3}$ to $\sigma_1 \approx 12.62$, $\sigma_2 \approx 2.21 \cdot 10^4$ and $\sigma_3 \approx 0.645$.
Inserting Eqs.\,(\ref{eq:rho_ld}) and (\ref{eq:rho_0}) into Eq.\,(\ref{eq:conversion}) yields the increase of the number of regular particles:
\begin{equation}
 \frac{\partial}{\partial t} N_{\textbf{r}}(t) = \frac{\gamma}{T} (1-\gamma)^{t/T} \text{erf}\left(\frac{X_d}{\sigma^0_t}\right).
\end{equation}
Hence, the growth of the number of regular particles, $\partial_t N_{\textbf{r}}(t)$, decreases in time due to two effects: first, because of the diffusive spreading of the ensemble which transports the particles out of the disorder region and is described by the monotonically decreasing error function and second, 
because of the fact that the number of particles which are available to be converted into $N_{\textbf{r}}(t)$ decreases with every particle that has already been converted, which is accounted for by 
the exponentially decaying term.

So far we have not distinguished between the different regular structures of the PSUS (Fig.\,\ref{fig:PSUS}) in which particles may be converted, i.e. $\gamma$ describes the conversion rate into any one of them. 
However, in order to get the desired particle density, we need to do so and distinguished between the different regular structures. 
Therefor, we introduce individual rates for any of the accessible regular components $\gamma_i$, where each $\gamma_i$ accounts for one of the ballistic islands
of the PSUS and two additional rates $\gamma^{\pm}_{\text{f}}$ account for the conversion into any of the regular curves above (below) the FISCs at positive (negative) velocities. 
Hence, the total rate, as used before, is given as 
\begin{equation}
 \gamma= \gamma^{+}_{\text{f}}+ \gamma^{-}_{\text{f}} + \sum_i \gamma_i.
\end{equation}
Thus, the change in the number of particles which are confined permanently within one of the ballistic islands is given by:
\begin{equation}
 \frac{\partial}{\partial t} N_{\textbf{r},i}(t) = \frac{\gamma_i}{T} (1-\gamma)^{t/T} \text{erf}\left(\frac{X_d}{\sigma^0_t}\right).
 \label{eq:conversion_sep}
\end{equation}
and equivalently for the particles above (below) the FISCs denoted by $N^{\pm}_{\textbf{r},f}(t)$. Note that for the exponentially decaying term, the total conversion rate $\gamma$
is still the relevant quantity, while the overall prefactors of $\gamma_i$ or $\gamma^{\pm}_{\text{f}}$ are related to the specifics of the considered regular structure.  
The advantage of this decomposition is, that all particles associated to one of the $N_{\textbf{r},i}(t)$ have the same average velocity $\bar v_i$, which is the velocity of the regular islands
central fixed point. Because of this, the spatial density, $\rho_i$, of particles within a given island can be calculated readily as:
\begin{equation}
\begin{aligned}
 \rho_{\textbf{r},i}(x,t)&=&\frac{\delta[x,\bar v_i]}{|\bar v_i|}\frac{\partial}{\partial t} N_{\textbf{r},i}\left( t- \frac{|x|-X_d}{|\bar v_i|}   \right), \\
 &=& \frac{\delta[x,\bar v_i]}{T |\bar v_i|} \gamma_i (1-\gamma)^{\tau/T} \text{erf}\left(\frac{X_d}{\sigma^0_{\tau}}\right)
  \label{eq:density}
\end{aligned}
\end{equation}
where $\tau \equiv t-(|x|-X_d)/|\bar v_i|$, $\delta[x,\bar v_i]$ equals one if $\text{sgn}(x)=\text{sgn}(\bar v_i)$ and zero otherwise and we set $\partial_t N_{\textbf{r},i}(t)=0$ for $t<0$.
For the particles above or below the FISCs the situation is more complicated since, here, every regular curve has a different velocity.  
However, we may use the average velocity $\bar v^{\pm}_f$ of the particles included in $N^{\pm}_{\textbf{r},f}(t)$, in which   
case Eq.\,(\ref{eq:density}) can be applied to obtain the corresponding densities $\rho^{\pm}_{\textbf{r},f}(x,t)$. The total density of regular particles is then given as:
\begin{equation}
 \rho_{\textbf{r}}(x,t) =  \rho^{+}_{\textbf{r},f}(x,t) + \rho^{-}_{\textbf{r},f}(x,t) + \sum_i \rho_{\textbf{r},i}(x,t).
 \label{eq:density_sum}
\end{equation}
In principle, the density of regular particles $\rho_{\textbf{r}}$ could be reproduced by means of Eqs.\,(\ref{eq:density}) and (\ref{eq:density_sum})
if one would treat the conversion rates $\gamma_i$ and $\gamma^{\pm}_f$ as fit parameters. However, in order to demonstrate that our modeling of the density evolution 
indeed describes the physical processes correctly, we extract all conversion rates directly from our numerical simulations and thus fix the values of all involved parameters.
Since our considered Hamiltonian is invariant under $x\rightarrow -x$, $p\rightarrow -p$,  
$t\rightarrow t+\pi/\omega$ we restrict ourselves to the regular structures with $\bar v<0$. 
The PSUS of our system features only a single notable ballistic island with a negative velocity (see Figs.\,\ref{fig:PSUS} and \ref{fig:rho_v} (e)). Hence the total density is decomposed in the following as: 
$\rho^{LD}(x<-X_D)=\rho_{\textbf{r},1} + \rho^{-}_{\textbf{r},f} + \rho_{\textbf{nr}}$.
In order to explicitly determine the conversion rates, we again consider the particle ensemble propagate in the setup containing localized disorder as before. 
At some fixed time $t=10^5$ we partition the ensemble into different parts depending on the particles coordinates in phase space (compare Fig.\,\ref{fig:rho_v} (e)) and then count the number of particles in each part: 
Out of the entire ensemble of $N=5\times 10^4$ particles, we find $N_1=9681$ particles at positions $x<X_d$ within the ballistic island which are thus described by $\rho_{\textbf{r},1}$ and 
below the FISC are $N_f^-=9940$ particles (belonging to $\rho^-_{\textbf{r},f}$).
Additionally, we get the total conversion rate $\gamma$ by 
using it as a fit parameter to the numerically obtain results of Eq.\,(\ref{eq:rho_ld}), yielding $\gamma\approx 10^{-5}$. By exploiting that $N_1/N_f^-= \gamma_1/ \gamma_f^-$ and that $\gamma_f^- + \gamma_1 = \gamma/2$
we get:
\begin{equation}
 \gamma_1= \frac{\gamma}{2(1+N_f^-/N_1)},\ \gamma_f^-= \frac{\gamma}{2(1+N_1/N_f^-)},
\end{equation}
yielding $\gamma_1 \sim 2.47 \times 10^{-6}$ and  $\gamma_f^-  \sim 2.53 \times 10^{-6}$. 
The velocity of the ballistic island as well as the average velocity of the particles below the FISC are determined as: $\bar v_1\approx -5.70$ and $\bar v^{-}_f \approx -6.51$.
Finally, we can compute the densities $\rho_{\textbf{r},1}$ and $\rho_{\textbf{r},f}^-$ according to Eq.\,(\ref{eq:density}) with no free parameter left. The resulting densities are shown in Fig.\,\ref{fig:rho_x} (d) and show very good agreement to the numerically obtained ones.
Note that the sudden drop to zero of $\rho_{\textbf{r},1}$ as seen both for the model as well as for the numerical data is simply at position $x=-(|\bar v_1|t + |X_d|)$, corresponding to the maximal distance 
that a particle can travel until time $t$ in a ballistic island with velocity $\bar v_1$. That this drop is smoothed out for $\rho_{\textbf{r},f}^-$, that is for particles below the FISC, is because, as mentioned before,
theses particles all have slightly different velocities and we used their average velocity $\bar v^{-}_f$ in the theoretical model.

%%%%%%%%%%%%%%%%%%%%%%%%%%%%%%%%%%%%%%%%%%%%%%%%%%%%%%%%%%%%%%%%%%%%%%%%%%%%%%%%%%%%%%%%%%%%%%%%%%%%%%%%%%%%%%%%%%%%%%%%%%%%%%%%%%%%%%%%%%%%%%%%%%%%%%%%%%%%%%%%%%%%%%%%%%%%%%%%%%%%%%%%%%%%%%%%%
\section{Transport in globally and locally disordered lattices.}
\label{S4}

A typical question of interest in the non-equilibrium dynamics of driven lattices, is whether or not a setup is transporting, that is whether the mean velocity of a particle ensemble takes some non zero 
asymptotic value. In fact, the unperturbed lattice as studied above constitutes one the paradigmatic example for a Hamiltonian ratchet setup and is known to yield 
directed transport for a biharmonic driving law of the form:
\begin{equation}
d_b(t)=A_1 \sin(\omega t) + A_2 \sin(2\omega t + \phi),
\label{eq:driving2}
\end{equation}
if $\phi\neq 0,\pi$ \cite{flach_directed_2000}.
Transport in these setups is usually studied for ensembles initialized at zero or small momenta and can be understood as a consequence of the asymmetrization of the systems chaotic layer around $v=0$ \cite{schanz_classical_2001, flach_directed_2000, denisov_tunable_2014}. 
From the analysis performed in the previous section, it is clear however that even if the ensemble is started at zero momentum, it will not be restricted to the PSUSs chaotic layer as soon as disorder is included.
Hence, it is an intriguing prospect to study the transport behaviour of the lattice in the presence of localized or global disorder, as e.g. 
the conversion into regular structures should lead to a violation of the sum rule of Hamiltonian ratchets \cite{schanz_classical_2001}.

%%%%%%%%%%%%%%%%%%%%%%% FIG 5
\begin{figure}[htbp]
\centering
\includegraphics[width=0.5\columnwidth]{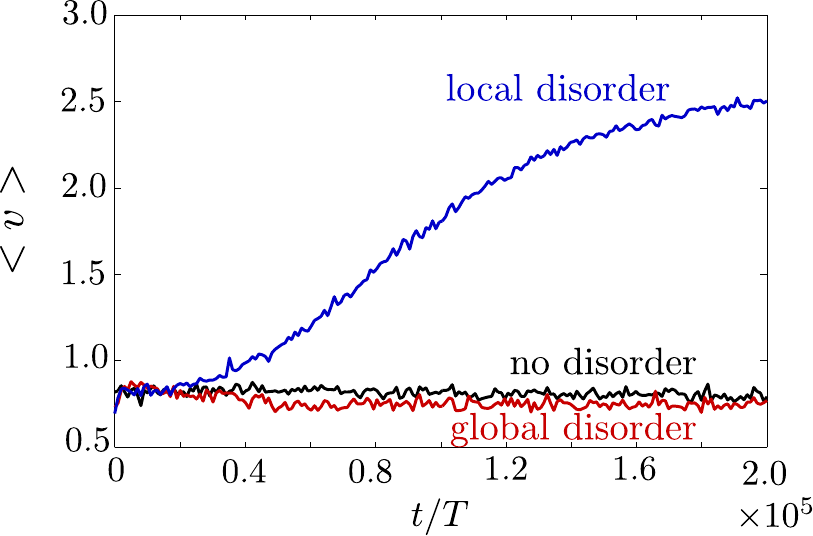}
\caption{\label{fig:transport} Mean velocities $<v>$ of particle ensembles for no disorder, global and local disorder as a function of time. Parameters of the lattice are the same as in Fig.\,\ref{fig:PSUS}. Parameters of the biharmonic driving law $d_b(t)$ are $A_1=1.0$, $A_2=0.5$, $\omega=3.0$ and $\phi=-7\pi/4$. For the local disorder case, the extension of the disorder domain was increased to $10^5 \times L$.}
\end{figure}
%%%%%%%%%%%%%%%%%%%%%%%%%%%%%%%%%%%%%%%%%%%%%%%%%%%%%
In order to determine the transport, we initialize the particle ensembles as before and determine their evolutions for the same setups but with the biharmonic driving law $d_b(t)$ (cf Eq.\,(\ref{eq:driving2})). The temporal evolution of the ensembles mean velocities are shown for the unperturbed lattice, as well as for localized and global disorder in Fig.\,\ref{fig:transport}. For the unperturbed case, we observe the usual behaviour of a Hamiltonian ratchet: very quickly (in fact on timescales hardly visible in Fig.\,\ref{fig:transport}), the ensemble fills the 
chaotic layer ergodically and the mean velocity fluctuates around the mean value of a trajectory in the chaotic layer, which we denote as $J^0$ in the following and which can be read off to be $\approx0.8$ for the used set of parameters. 
For the globally disordered lattice, the mean velocity 
saturates against an asymptotic value of $J^\text{GD}$ which differs only slightly from its disorder free counter part $J^0$. In strong contrast, the local disorder leads to a significant increase of the mean velocity, which 
only slowly saturates towards its asymptotic value of $J^\text{LD} \approx 2.5$. 
In order to make this effect more apparent, we have increased the extension of the disorder domain as compared to the previous sections by a factor of $10$ to $10^5\times L$.

According to the sum rule of Hamiltonian ratchet transport, the asymptotic mean velocity of an ensemble initialized in the chaotic layer is given by the difference of the mean velocities of the FISCs at positive and negative velocities, minus the mean velocities of all ballistic islands weighted by their area in the systems PSS \cite{schanz_classical_2001}. In case of global disorder, the ballistic islands as well as the regular curves of the PSUS at higher energies become populated and thus may add to the directed transport. This process however, appears equally for regular structures at positive and negative velocities and thus the overall transport is not expected to change significantly. In this sense, one might say that the global disorder does not add to the desymmetrization of the system and hence has only little influence on the systems transport properties. 
For the localized disorder, this line of arguments holds as long as most particles are still located within the disorder region. Indeed we find, that $<v>$ follows the other two cases at early times $t/T\lesssim0.2 \cdot 10^5$ and deviates strongly only afterwards. The strong increase of the transport at later times can be understood straightforwardly by means of the analysis as performed before: At the edges of the disorder domain at $x= \pm X_D$, the ensemble is injected into the regular structures of the PSUS at some rates $\gamma_i$ or $\gamma_f^{\pm}$ and the number of particles within a given regular structure at a given time is assumed to be proportional
to the number of particles which have crossed $x=\pm X_D$.
In the previous section however, the ensemble was modeled simply as a Gaussian with some diffusive spreading described by its time dependent width $\sigma_t$. In a transporting systems, this description is insufficient as it neglects the overall drift of the ensemble within the disorder domain with a positive velocity of $J^\text{GD} \approx J^0$. Hence, more particles pass $x=+X_D$ per time as compared to $x=-X_D$ and thus conversion into regular structures with a positive velocity is enhanced. 
After some time, most particles are either converted permanently into regular structures of the PSUS or are transported diffusively out of the disorder domain and the mean velocity saturates against its asymptotic value $J^\text{LD}$. In analogy to the density (Eq.\,(\ref{eq:density_sum})), the transport can be written as:
\begin{equation}
J^\text{LD}= n_{\textbf{nr}} J^0 + n^{+}_{\textbf{r},f} \bar v^{+}_f
+ n^{-}_{\textbf{r},f} \bar v^{-}_f +  \sum_i  n_{\textbf{r},i}\bar v_i,
\end{equation}
where the $n's$ denote the asymptotic ratios of the ensemble belonging to the respective part of the ensemble decomposition as introduced in section \ref{S3}: $n_{\textbf{r},i}\equiv \lim_{t\rightarrow \infty}N_{\textbf{r},i}(t)/N$
and equivalently for particles above or below the FISCs.
Note that by setting the transport velocity of the part of the ensemble which is not converted permanently to a regular structure to $J^0$, we have assumed that asymptotically all particles will have left the disorder domain and are thus governed by the PSUS with associated current $J^0$.

%%%%%%%%%%%%%%%%%%%%%%%%%%%%%%%%%%%%%%%%%%%%%%%%%%%%%%%%%%%%%%%%%%%%%%%%%%%
\section{Conclusions.}
\label{S5}

We have studied how the dynamics of classical particles in a Hamiltonian driven lattice potential is altered by the inclusion of disorder. 
If the disorder is localized within a finite domain of the lattice, the particles are accumulated into the regular structures of the phase space of the unperturbed system. For this case, we show how the particle density 
inherits strongly pronounced tails, which at late times, even lead to an increase of the density towards larger positions.  Hence, the localized disorder leads to a strong modification of the usual Gaussian like diffusion as observed in the unperturbed lattice.
The observed particle densities are explained by means of a decomposition of the entire ensemble into regular and diffusive parts, where the regular part is decomposed further as a sum over all regular structures of the unperturbed phase space. We derive an approximate equation for the densities tail which accounts for the conversion from the diffusive part of the ensemble
into the regular ones. All parameters of the resulting model can be fixed by numerical simulations, leading to a derived spatial dependence of the density with no free fit parameters left and which is shown to be in very good agreement with the full simulation.
  
Finally, we study the transport properties of a disordered lattice. While for global disorder, the asymptotic transport velocity is modified only slightly compared to the unperturbed lattice, we demonstrate that transport is increased significantly by the localized disorder.
Here, we propose an extension to the usual sum rule of Hamiltonian ratchet transport which incorporates the effect of 
conversion processes from diffusive to regular parts of the unperturbed phase space as made possible by the inclusion of disorder.

%%%%%%%%%%%%%%%%%%%%%%%%%%%%%%%%%%%%%%%

\bibliography{bibliography}{}

\end{document}